# Investigating the free-floating planet mass by Euclid observations

Lindita Hamolli[1] • Mimoza Hafizi[1] •
Francesco De Paolis[2,3] • Achille A. Nucita[2,3]

1- Department of Physics, University of Tirana, Albania.
2- Department of Mathematics and Physics "Ennio De Giorgi", University of Salento, CP 193, I-73100 Lecce, Italy.
3- INFN, Unit of Lecce, University of Salento, CP 193, I-73100 Lecce, Italy.

**Abstract.**
The detection of anomalies in gravitational microlensing events is nowadays one of the main goals among the microlensing community. In the case of single-lens events, these anomalies can be caused by the finite source effects, that is when the source disk size is not negligible, and by the Earth rotation around the Sun (the so-called parallax effect). The finite source and parallax effects may help to define the mass of the lens, uniquely. Free-floating planets (FFPs) are extremely dim objects, and gravitational microlensing provides at present the exclusive method to investigate these bodies. In this work, making use of a synthetic population algorithm, we study the possibility of detecting the finite source and parallax effects in simulated microlensing events caused by FFPs towards the Galactic bulge, taking into consideration the capabilities of the space-based Euclid telescope. We find a significant efficiency for detecting the parallax effect in microlensing events with detectable finite source effect, that turns out to be about 51% for mass function index $\alpha_{PL} = 1.3$.

**Keywords** Free-floating planets; gravitational microlensing

## 1. Introduction

A microlensing event occurs when a compact massive object (lens) approaches very closely to the observer's line of sight toward a background star (source). The standard microlensing light curve follows a characteristic symmetric curve, called Paczynski profile, which is based on the assumptions that (i) the source star and the lens are point-like and (ii) the relative motion among the observer, lens and source are uniform and linear (Paczynski 1986). The first assumption is often not valid, particularly in very bright microlensing events. Indeed, in these events the finite source effects are often non negligible (Witt & Mao 1994). Furthermore, stars with their bound planets (binary lenses) represent systems far from being point-like lenses. A planet with mass in the range $[1 \div 10] M_\oplus$ can cause deviations with respect to the standard single lens light curve with amplitudes larger than 10% and lasting for a few hours or more (Bennett 1996). It is a matter of fact that these deviations are a precious source for discovering extrasolar planets. Nowadays, most of the extrasolar planets detected by microlensing are situated inside the "lensing zone", i.e. at distances since about 0.5 AU to 8 AU from their host star (Han 2009). Assumption (ii) is clearly wrong in principle, as we know the Earth revolves around the Sun and thus the Earth acceleration perturbs the microlensing light curves and may give rise to noticeable asymmetry (Alcock et al. 1995).

In recent years astronomers are discovering in the Milky Way planetary objects with mass $M \leq 0.01 M_\odot$ that are not bound to a host star (see e.g. Zapatero Osorio et al. 2000) or are on orbit with semi-major axis larger than about 100 AU. These objects are called free-floating planets, FFPs, (see e.g. Strigary et al. 2012). Since they are quite dark objects, it is very hard to directly detect them at distances larger than a few tens of parsecs. Up to date, the only method to discover and investigate the FFP population is through gravitational microlensing. And indeed, recently, gravitational microlensing has allowed the detection of ten FFPs towards the Galactic bulge by ground-surveys (Sumi et al. 2011). However, microlensing does not allow to determine uniquely the FFP mass, due to the parameter degeneracy problem.

The situation might significantly change with the advent of space-based surveys with very high monitoring sampling capabilities, which will help to detect many short-duration microlensing events caused by FFPs. Actually, several space-based experiments have been proposed, such as observations by the Euclid telescope (Laureijs et al. 2011). Euclid is scheduled to be launched in 2020 at the L2 Langrange point, and, in addition to scan about 1/3 of the sky for mapping the dark matter and dark energy distribution with unprecedented sensitivity, it will likely be able to perform also microlensing observations towards the Galactic bulge.

Here, continuing on the line of our previous work (Hamolli et al. 2013(a) and Hamolli et al. 2015), we focus on the investigation of the parallax and the finite source traces, left on microlensing event light curves, with the aim of obtaining a realistic treatment of the expected events observable by Euclid. In our calculation, the FFPs are considered as free, unbound objects, giving rise to Paczynski light curves. Indeed, were they be bound to a host star, they would lie at very far distances from it, far way from the "lensing zone", making them undetectable in the light curve eventually produced by their host star in a microlensing event. Moreover, since FFPs should have been formed in proto-planetary disks and subsequently scattered into unbound or very distant orbits, we assume that their spatial and velocity distributions are the same as those of the stars in our Galaxy.

The paper is organized as follows. In section 2, we describe the methodology to measure the Einstein radius projected onto the observer plane $r_E$ and the angular Einstein radius $\theta_E$, by which we can determine the mass of the lens. In section 3, the results of our Monte Carlo simulations are illustrated, while some conclusions are briefly drawn in section 4.

**2. Determination of the Lens Mass**

A Paczynski microlensing light curve depends on three unknown parameters (Paczynski 1986): the time of maximum amplification $t_0$, the Einstein time $T_E = R_E/v_T$ (where $R_E$ is the Einstein radius and $v_T$ is the lens transverse velocity) and the impact parameter $u_0$ (the minimum value of the separation $u(t)$ between the lens and the line of sight in $R_E$ units). However, among these parameters, only $T_E$ contains information about the lens. This gives rise to the so called parameter degeneracy problem. To the aim of breaking this degeneracy we consider the finite source and parallax effects induced in the microlensing light curves by the source size and Earth orbital motion around the Sun.

**2.1 Parallax Effect**
The parallax effect, due to the motion of the Earth around the Sun, may leave in microlensing events observable features (Alcock et al. 1995), which can be used to constrain the lens parameter. Following Dominik (1998), the trajectory of the Earth (observer) projected onto the lens plan can be described by the coordinates:

$$x_1(t) = \rho\{-\sin\chi\cos\phi(\cos\xi(t)-\varepsilon) - \sin\chi\sin\phi\sqrt{1-\varepsilon^2}\sin\xi(t)\}$$
$$x_2(t) = \rho\{-\sin\phi(\cos\xi(t)-\varepsilon) + \cos\phi\sqrt{1-\varepsilon^2}\sin\xi(t)\}$$

where $\rho = \dfrac{a_\oplus(1-x)}{R_E} = \dfrac{a_\oplus(1-x)}{r_E(1-x)} = \dfrac{a_\oplus}{r_E}$ is the length of the Earth orbit semi-major axis projected onto the lens plane and measured in Einstein radii. Here, $a_\oplus$ is the semi-major axis of the Earth orbit around the Sun, $\varepsilon = 0.0167$ is the Earth orbit eccentricity and $x = D_l/D_s$, where $D_s$ and $D_l$ are the source-observer and lens-observer distance, respectively.

The position of the source star is characterized by the parameters $\phi, \chi$ which give, respectively, the longitude measured in the ecliptic plane from the perihelion towards the Earth motion and the latitude measured from the ecliptic plane towards the northern point of the ecliptic.

By fitting the event light curve with the model described above, one can determine the parameter $\rho$. Since $a_\oplus$ is known, one can find the Einstein radius projected onto the observer plane, $r_E$. In fact, by ground-based observations, the $r_E$ value can be defined for the sub-class of relatively long events (with duration about a few months). Alcock et al. (1995) have presented the first detection of parallax effects in a gravitational microlensing event. Their description of the parallax effect in the light curves was obtained by expanding the Earth trajectory up to the first order in the eccentricity $\varepsilon$ (Dominik 1998).

However, in the case of space observatories (like Euclid positioned at the L2 point) the satellite acceleration around the Sun produces a parallax effect that can be detected also in short-duration microlensing event light curves ($\leq 1-2$ days, Hamolli et al. 2013(a)).

**2.2 Finite Source Effect**
As already discussed by Hamolli et al. (2015) another way to break the parameter degeneracy is through the finite source effect, which is more easily detectable in bright microlensing events. In these events the value of $u_0$ becomes comparable to the source radius projected onto the lens plane in units of the Einstein radius, $\rho_*$ and the resulting microlensing light curve deviates from the standard form of a point-source event (Han et al. 2004). These deviations depend on the light intensity distribution throughout the source stellar disk. Different brightness profiles have been proposed and discussed in the literature. Among them, the one that describes the light intensity distribution in the stellar disk more accurately than any other model is the non-linear limb-darkening model (Claret 2000), which is adopted in the following discussion.
This model is described by the equation:

$$I(\mu) = I(1)\left[1 - a_1\left(1 - \mu^{1/2}\right) - a_2\left(1 - \mu\right) - a_3\left(1 - \mu^{3/2}\right) - a_4\left(1 - \mu^2\right)\right]$$

where $\mu = \cos(\theta)$, $\theta$ is the angle between the line of sight and the emergent intensity and $a_1$, $a_2$, $a_3$ and $a_4$ are the limb darkening coefficients (LDCs).

By fitting the microlensing light curve with the Claret model for the source's disk limb darkening profile, one can define the parameter $\rho_*$. If the angular size, $\theta_*$ of the source may be estimated through the color and the absolute magnitude of the source and since the parameter $\rho_*$ is the normalized source angular radius in units of $\theta_E$, the angular Einstein radius can be measured by $\theta_E = \theta_*/\rho_*$ (Zub et al. 2011).
For example, Zub et al. (2011) have presented a detailed analysis of the highly sampled OGLE 2004-BLG-482 event and have determined the source star LDCs and the angular Einstein radius, which results to be $\theta_E = \theta_*/\rho_* \simeq 0.4$ mas.

Once $\theta_E$ and $r_E$ are measured, the lens mass is uniquely determined (Jiang et al. 2004) through the relation

$$M = \frac{c^2}{4G} r_E \theta_E \tag{1}$$

## 3. Monte Carlo Calculations and Results

We investigate the possibility of parallax and finite source effects in observations by Euclid, using Monte Carlo numerical simulations of microlensing events towards its field of view. For each event we extract, via the procedure detailed below, nine parameters: FFP distance, FFP mass, FFP transverse velocity, event impact parameter, source distance and the four source LDCs. In particular:
1) We generate the FFP distance from the observer, $D_l$, based on the disk and bulge FFP spatial distributions, which is assumed to follow that of the stars (Gilmore et al. 1989, De Paolis et al. 2001, Hafizi et al. 2004) along the Euclid line of sight to the Galactic bulge. Since the Euclid galactic coordinates are $b = -1.7°$, $l = 1.1°$, by the usual transformation relations between coordinate systems we find the following values for parameters: $\phi = 167.8°$ and $\chi = -5.4°$.
2) We generate the lens mass randomly extracting $M$ from a distribution probability following the mass function defined by Sumi et. Al (2011), i.e. $\frac{dN}{dM} = k_{PL} M^{-\alpha_{PL}}$, with mass function index $\alpha_{PL}$ in the range [0.9-1.6].
3) We extract the FFP relative transverse velocity assuming for each coordinate the Maxwellian distribution

$$f(v_i) \propto e^{-\frac{(v_i - \bar{v}_i)^2}{2\sigma_i^2}}, \qquad i \in \{x, y, z\}$$

where the coordinates (x, y, z) have their origin at the galactic center and the x and z-axes point to the Sun and toward the north Galactic pole, respectively. Here we mention that we are interested only to the velocity components perpendicular with respect to the line of sight, namely to the y and z components. For lenses in the Galactic bulge we use the mean velocity components $\bar{v}_y = \bar{v}_z = 0$, with dispersion $\sigma_y = \sigma_z = 100$ km/s; for lenses in the Galactic disk we use the mean velocity components $\bar{v}_y = 220$ km/s, $\bar{v}_z = 0$, with dispersion velocity $\sigma_y = \sigma_z = 30$ km/s for the thin disk and $\sigma_y = \sigma_z = 50$ km/s for the thick disk (Han 1995).

4) The event impact parameter $u_0$ is randomly extracted from a probability distribution uniformly distributed in the interval [0, 6.54]. The upper value of that interval follows from the observation that the amplification threshold for a detectable Paczynski microlensing event from space-based observations is $A_{th} = 1.001$.

5) We extract the source distance $D_s$ from a distribution probability following the Galactic bulge spatial distributions of the stars: $f(x, y, z) \sim e^{-s^2/2}$, with $s^4 = (x^2/a^2 + y^2/b^2)^2 + z^4/c^4$, $a = 1.49$ kpc, $b = 0.58$ kpc and $c = 0.40$ kpc (for further details see Hafizi et al. 2004).

6) We generate a synthetic stellar population of 10000 stars with luminosity, effective temperature, gravity acceleration and absolute magnitude using the free-available code (Aparicio & Gallart, 2004) developed at the Instituto de Astrofisica de Canarias (IAC, http://iac-star.iac.es). We then consider a sub-sample of 1000 randomly extracted stars by using and adapted version of the HIT or MISS algorithm (see e.g. Gentile 1998). In particular, we extract two random variables ($x_i$ and $y_i$) uniformly distributed between 0 and 1. Then we evaluate the corresponding log $g_i$ and log $T_{eff,i}$ values via a) log $g_i$ = log $g_{max}$+( log $g_{max}$- log $g_{min}$)$x_i$ and b) log $T_{eff,i}$ = log $T_{eff,max}$+( log $T_{eff,min}$ - log $T_{eff,min}$)$y_i$.

In this case, the resulting values are still uniformly distributed between the corresponding minimum and maximum values. We then extract $x_{i+1}$ (again uniformly distributed between 0 and 1) and the acceptable values of log $g_i$ and log $T_{eff,i}$ are selected requiring that c) $x_{i+1}$ max(f(log $g_i$, log $T_{eff,i}$)) ≤ f(log $g_i$, log $T_{eff,i}$ ), where f(log $g_i$, log $T_{eff,i}$) is a two dimensional distribution obtained by sampling with an opportunely spaced grid the synthetic stellar population. This method ensures that the obtained couples of (log $g_i$, log $T_{eff,i}$) values follow the right distribution and finally, through the website: http://vizier.u-strasbg.fr/viz-bin/VizieR-3?source=J/A%2bA/363/1081/phoenix (Claret 2000) we obtain the source star LDCs: $a_1$, $a_2$, $a_3$ and $a_4$.

Taking in mind the performances of the Euclid telescope and the fact that the efficiency of parallax effect in microlensing light curves would depend on the time of the year in which the observations are performed (Hamolli et al. 2013(b)), that is on the value of parameter $\xi_0$ describing the satellite position at the event peak time $t_0$, we consider in particular two cases, i.e. when the Earth (and therefore also the satellite) are near the summer solstice ($\xi_0 = 165°$) and the equinox ($\xi_0 = 75°$). We also assume a sampling of 20 min and that Euclid will be located at the L2 point, with $a_\oplus = 1.01 AU$.

For each simulated event we calculate the standard light curve, the curve with finite source effects, the curve with parallax effects and the curve where the motion of the Euclid around the Sun is superposed over the finite source curve.
We assume that a microlensing event is detectable if in its light curve there are at least eight consecutive points with amplification larger than the Euclid threshold amplification $A_{th} = 1.001$. We also assume that the finite source and/or parallax effects can be detectable on a light curve when there are at least eight consecutive points with residuals $\text{Re} s > 0.001$ (between the curve with finite source or parallax effect and the standard curve, see Hamolli et al. 2013(a)).
The efficiency of the finite source and parallax effects is calculated as the ratio of the number of events with detectable finite source and parallax effects to the total number of detectable events. We have considered separately the FFP distribution in the Galactic bulge, thin disk and thick disk and defined the efficiency with respect to the value of $\alpha_{PL}$ (ranging from 0.9 to 1.6).

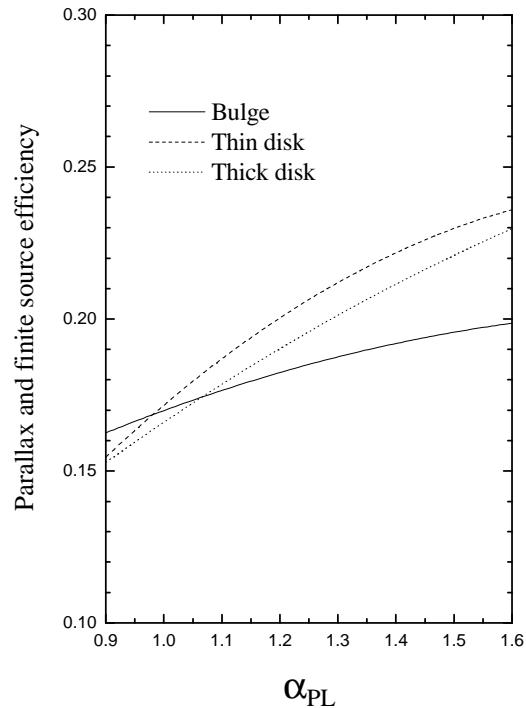

**Fig. 1** The efficiency of the parallax and finite source effects in microlensing events caused by FFPs as a function of $\alpha_{PL}$ for the three different distributions of FFPs: bulge (black line), thin disk (dashed line) and thick disk (dotted line) near the summer solstice.

In Figure 1 we show the obtained results for the efficiency of the parallax and finite source effects. As one can see, the efficiency values tend to get higher with increasing the $\alpha_{PL}$ values. The single efficiencies are shown, respectively, in Fig. 3 and Fig.4 in the papers of Hamolli et al. 2013(a), and Hamolli et al. 2015. In particular, the parallax efficiency values decrease with increasing the $\alpha_{PL}$ values, while the finite source efficiency tends to become larger with increasing $\alpha_{PL}$.

We have also calculated the detection efficiency of FFP events with parallax effect with respect to the events with detectable finite source effects. To this aim we calculate the residuals between the light curve containing both finite source and parallax effects and the light curve containing only the finite source effect. The ratio of the number of the finite source events with detectable parallax effects to the number of detectable finite source events gives the efficiency of parallax effect in events with detectable finite source effect.

In Figure 2 we show the obtained results for FFPs in bulge, thin disk and thick disk with respect to the value of $\alpha_{PL}$.

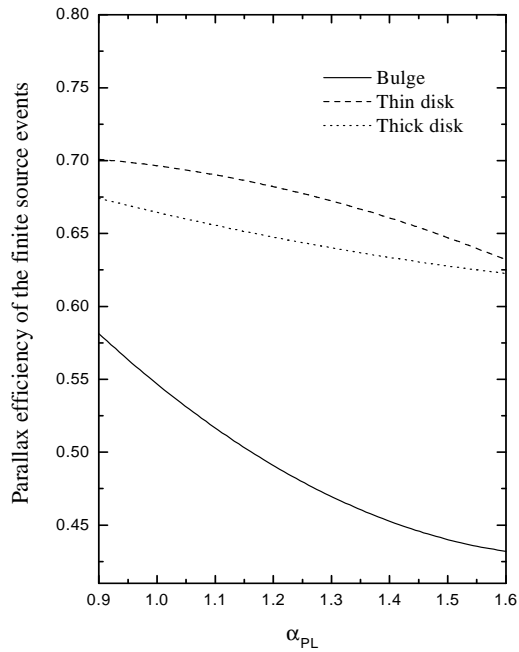

**Fig. 2** The parallax efficiency in FFP events with detectable finite source effects, as a function of $\alpha_{PL}$ for the three different distributions of FFPs: bulge (continued line), thin disk (dashed line) and thick disk (dotted line) near the summer solstice.

As one can see, the efficiency of parallax effect in events with detectable finite source effect is larger for the FFPs in the thin disk, while it is smaller for FFPs populating the thick Galactic disk and the bulge, but it always tends to decrease with increasing the $\alpha_{PL}$ values.

We then consider the efficiency assuming that the Euclid satellite lies near the equinox at the moment of the event peak time (i.e. $\xi_0 = 75°$). In this case we find that the parallax efficiency of the finite source events for $\alpha_{PL} = 1.3$ decreases significantly, approximately of a factor 3.7 in the bulge, 1.9 times in the thin disk and 2.2 times in the thick disk.

**4. Conclusions**
In this paper we have considered the deviations due to finite source and parallax effects in the standard microlensing light curves towards the Galactic bulge caused by FFPs taking into account the observing capabilities of the future Euclid telescope.

In our calculations it is assumed that the source star disks have circular shape on the sky and that the limb-darkening profile is given by the four-parameter non-linear power law model (Claret 2000).

Using Monte Carlo numerical simulations, we investigate the efficiency to detect finite source effects and parallax effects in microlensing light curves caused FFPs.

In a previous paper (Hamolli et al. 2013(a)) it was found that Euclid observations towards the Galactic bulge should allow to detect a substantial number of events caused by bulge FFPs and a minor number of events caused by FFPs belonging to the thin and thick disks populations. Here we find that in the case of the middle value of $\alpha_{PL} = 1.3$ about 37% of all detectable events show their finite source effects and among them about 51% also manifest the parallax effect. So, we have that about 19% of all detectable events are expected to show both effects and consequently give the possibility to identify unambiguously the FFP mass.

Finally, with the aim of best planning the time of Euclid's microlensing observations, we also note that the best period during the year for detecting microlensing events due to FFPs manifesting both the finite source and the Earth parallax effects is around the summer solstice. For these events the lens mass may be determined with a greater efficiency.

**Acknowledgments**
FDP and AAN acknowledge the support by the INFN project TAsP.

**References**

Alcock, C. et al., 1995, Astrophys. J., **454**, 125
Aparicio, A. & Gallart, C., 2004, Astrophys. J. **128**, 1465
Bennett D. P. & Rhie S. H. 1996, Astrophys. J. **472**, 660
Claret, A. 2000, Astron. Astrophys., **363**, 1081
De Paolis, F. et al., 2001, Astron. Astrophys., **366**, 1065
Dominik, M. 1998, Astron. Astrophys., **329**, 361
Gentile, J.E. 1998, Random Number generation and Monte Carlo methods, Springer, New York, ISBN 978-1-4757-2960-3

Gilmore, G. et al., 1989, Astron. Astrophys., **27**, 555
Hafizi, M. et al., 2004, Int. Journ. Mod. Phys. D, **13**, 1831
Hamolli, L. et al., 2013(a), Int. Journ. Mod. Phys. D, **22**, 1350072
Hamolli, L. et al., 2013 (b), Bulgarian Astronomical Journal, **19**, 34
Hamolli, L. et al., 2015, Advanced in Astronomy, **Article ID 402303**
Han, Ch & Gould, 1995 Astron. Astrophys., **447**, 53
Han, Ch. et al., 2004, Astrophys. J., **604**, 372
Han, Ch. et al., 2009, Astrophys. J., **691**, 452
Jiang, G. et al., 2004, Astrophys. J., **617**, 1307
Laureijs, R. et al., 2011, Preprint arXiv:**1110.3193**
Paczynski, B. 1986, Astrophys. J., **384**, 1
Strigary, L.E. et al., 2012, Mon. Not. R. Astron. Soc., **423**, 1856
Sumi, T. et al., 2011, Nature, **473**, 349
Witt, H.J. & Mao, Sh. 1994, Astrophys. J., **430**, 505
Zapatero Osorio, M.R. et al., 2000 Science, **290**, 103
Zub, M. et al., 2011, Astron. Astrophys., **525**, 15